\documentstyle[11pt,paspconf,epsf,psfig,twoside]{article}

\begin{document}
\noindent
\raisebox{60pt}[0pt][-60pt]{\parbox{\textwidth}
{\it Proc. of the Annapolis Workshop on Magnetic Cataclysmic
Variables, 1998 } }


\title{RX\,J0501.7--0359: a new ROSAT discovered 
eclipsing polar in the period gap}


\author{
Vadim Burwitz$^{1,2}$,
Klaus Reinsch$^{2}$,
Klaus Beuermann$^{2,1}$,
and Hans-Christoph Thomas$^{3}$
}
\affil{
$^{1}$Max-Planck-Institut f\"ur extraterrestrische Physik, Garching, Germany
$^{2}$Universit\"ats Sternwarte G\"ottingen, Germany\\
$^{3}$Max-Planck-Institut f\"ur Astrophysik, Garching, Germany
}

%

%


\begin{abstract}
RX\,J0501.7--0359 is a new $V\sim 17$\,mag eclipsing polar discovered 
during the ROSAT all-sky survey (RASS).  
Extensive follow-up observations at X-ray (pointed ROSAT PSPC and HRI
observations) and optical (time-resolved photometry and spectroscopy) wavelengths
were obtained which allowed us to determine a very precise orbital period and to 
provide constraints for the masses of the binary components and the orbital 
inclination. With an orbital period of $P \sim 171$\,min the system is placed at the
upper end of the period gap. 
From the radial velocity amplitude of the secondary and the duration of the primary 
eclipse we obtain a mass ratio $q = M_{2}/M_{1} = 0.83 \pm 0.15$ and an  
inclination $i = 75\deg \pm 3\deg$ for the system. This yields a white dwarf mass 
$M_{1} = 0.43^{+0.10}_{-0.07}\,M_{\sun}$. Cyclotron humps might be present in 
some of our optical spectra leading to a tentative magnetic field strength of 
$B\sim 25$\,MG.

\end{abstract}


\section{Introduction}
\noindent The ROSAT All-Sky Survey (RASS) revealed a great wealth of new objects
with soft X-ray spectra. Our optical identification of a sample of these
sources has led to a significant increase in the number of known polars 
(Beuermann 1998, Thomas et al. 1998). Detailed
X-ray and optical follow-up studies allow us to determine
their orbital periods, magnetic field strengths, accretion geometry, etc.
This leads to a largely increased data base of known polars which
is important for studying the general
properties of these systems as well as their evolution. Apart from this
it is important to study individual objects in great
detail as many of the new systems have remarkable characteristics such as 
extremely short periods and high magnetic field strengths 
(see Burwitz et al. 1997, 1998).

\section{Observations and discussion}

RX\,J0501.7-0359 (= 1RXS\,J050146.2--035927) was detected as a soft 
(hardness ratio $HR1 = (-0.96\pm 0.03$), bright ($0.22\pm 0.03$ cts/s), and  
variable X-ray source in the RASS (Voges et al. 1996).  
As its optical counterpart, we have identified an eclipsing
$V\sim 17$\,mag polar located at RA = $05^{\rm h}01^{\rm m}46\fs 1$, 
DEC = $-03\deg 59\arcmin 32\arcsec$.

\begin{figure}
\begin{center} 
\begin{minipage}{130mm}
\psfig{file=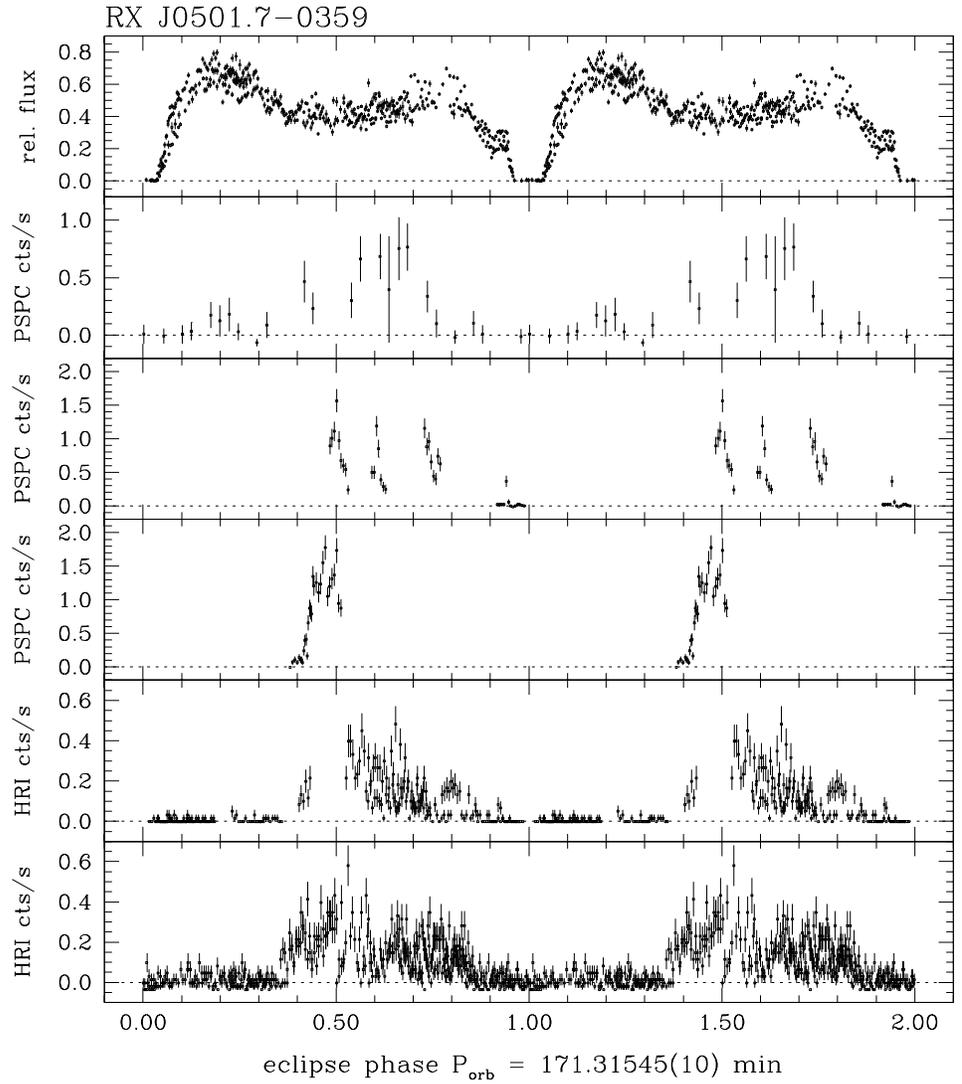,width=130mm}
\end{minipage}
\end{center}
\caption{\label{licu} Optical and X-ray
light curves of RX\,J0501.7--0359 folded over its 171 min orbital 
period using the ephemeris given in Eq.~1. 
From top to bottom: V band optical light curve (Jan. 11--12, 1996), 
ROSAT all-sky survey light curve (Aug. 24--26, 1990),
two short pointed ROSAT PSPC light curves (2.4\,ksec, Feb 24, 1992. and 1.8\,ksec,
Feb. 15--22, 1993 ), and two pointed ROSAT HRI light curves (16.5\,ksec, 
Sep. 8--16, 1995 and 28.8\,ksec Feb. 26 -- Mar. 19, 1996).}
\end{figure}
Here, we present the analysis of our large amount of follow-up observations of this new 
object: optical V-band CCD photometry with the Dutch 0.9-m telescope and time-resolved
spectroscopy with EFOSC2 at the ESO/MPI 2.2-m telescope on La Silla, Chile,
infrared J, H, and K photometry with MAGIC at the 3.5-m telescope on Calar Alto, Spain,
and X-ray data with ROSAT using both the PSPC and the HRI.


\begin{figure}
\begin{center} 
\begin{minipage}{120mm}
\psfig{file=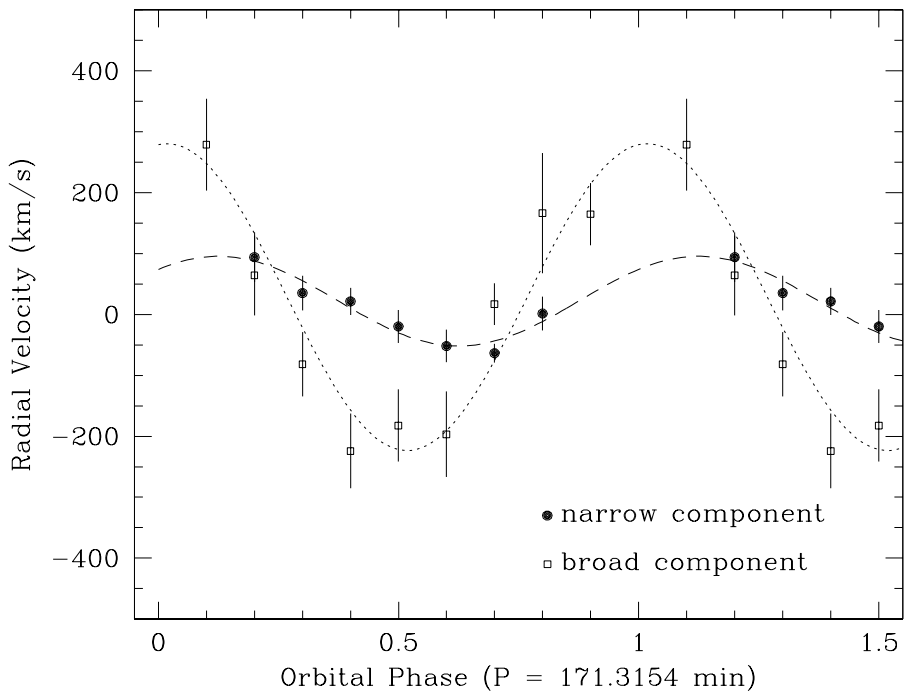,width=120mm}
\end{minipage}
\end{center}
\caption{\label{rvcurve}
Average radial velocity curves of the broad and narrow components observed
in the He{\sc II}\,4686\AA\ and H$\beta$ emission lines of RX\,J0501.7--0359.
}
\end{figure}
\begin{figure}[t]
\begin{center} 
\begin{minipage}{95mm}
\psfig{file=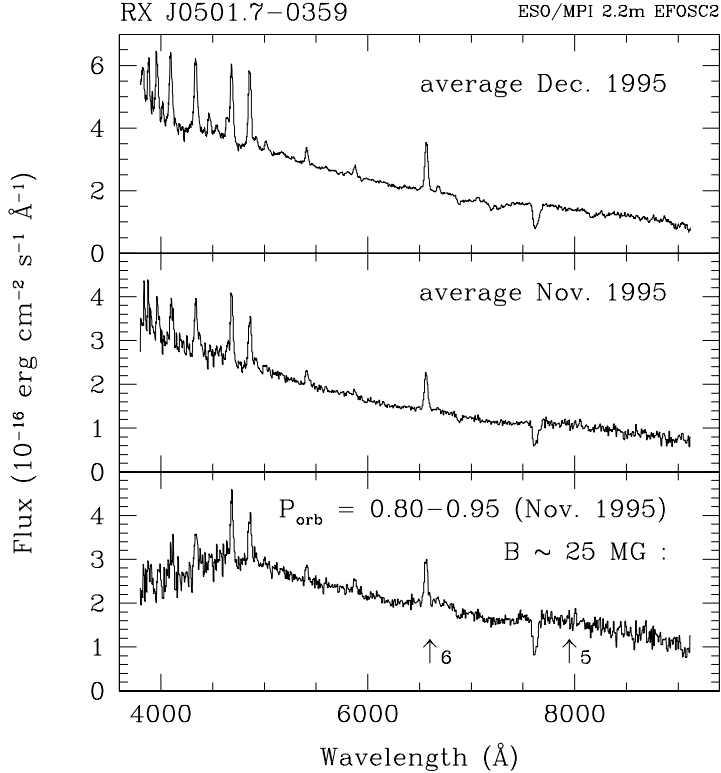,width=95mm}
\end{minipage}
\end{center}
\caption{\label{ospec} Average optical spectra obtained during different observing runs: 
(top panel) December 1995, orbital average,
(middle panel) November 1995, orbital average,
(bottom panel) Nov. 1995, orbital phase interval 0.80 and 0.95. 
The positions of two weak humps in the latter spectrum which we have tentatively 
identified as the 5th and 6th cyclotron harmonics corresponding to a 
magnetic field strength of B\,$\sim$\,25\,MG are indicated by arrows.} 
\end{figure}
The optical light curves show a strong modulation which could be caused either by
cyclotron beaming or by variations of the effective emitting area of the accretion 
stream as seen from 
different angles during the orbit. The light curve also features a deep total eclipse 
of the accreting white dwarf and the stream by the secondary star and a pre-eclipse dip, 
possibly
due to the accretion stream crossing our line-of-sight towards the accretion region 
(cf. Fig.~\ref{licu}, top panel).

From our optical, IR, and spectrophotometry, we have derived 16 mid-eclipse timings 
between August 1993 and January 1996. These were used to determine the orbital period 
of RX\,J0501.7--0359 very precisely and lead to the following ephemeris:
\begin{equation}
\label{eq.ephem}
T_{\rm mid-eclipse}({\rm HJD}) = 2449748.83782(20)+0.11896906(7)\times E.
\end{equation}

From the optical spectra we have determined radial velocities of the narrow 
and broad components of the Balmer (H$\alpha$, H$\beta$) and the He{\sc II}\,4686\AA \  
emission lines. The average radial velocity curve yields an 
amplitude K$'\!_{2}=(73.6\pm 25)$\,km/s for the narrow component (cf. Fig. \ref{rvcurve}).
Maximum redshifts of the broad and narrow components occur at orbital phases
0.02 and 0.12, respectively.

\begin{figure}
\begin{center} 
\begin{minipage}{100mm}
\psfig{file=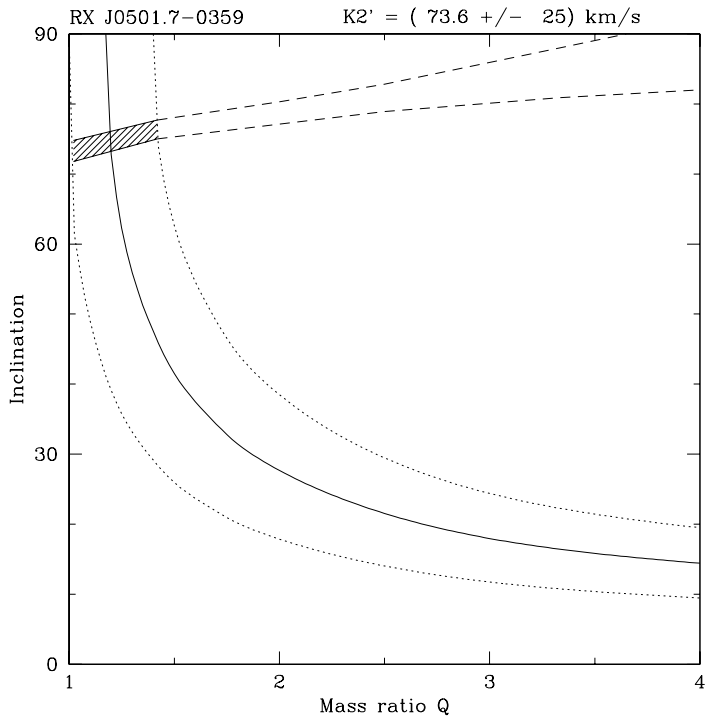,width=100mm}
\end{minipage}
\end{center}
\caption{\label{masses} Mass function (continuous line with dotted error limits) and 
eclipse duration constraint (dashed lines) for RX\,J0501.7--0359. The valid range for
the orbital inclination $i$ and the mass ratio $Q = M_1/M_2$ is restricted to the shaded 
area. For the mass-function, the radial velocity amplitude of the narrow emission line 
component has been corrected for the offset between the center-of-light and the 
center-of-mass of the illuminated secondary (see Beuermann \& Thomas 1990).
}
\end{figure}

Assuming a Roche-lobe filling secondary star, Kepler's laws define a relation between
the orbital period $P$ of a CV and the mass $M_2$ and radius $R_2$ of the 
secondary (Eq. 1 in Beuermann et al. 1998). As RX\,J0501.7-0359 is probably a system 
which has been born in the period gap, its secondary cannot be much evolved. Therefore, 
the theoretical mass-radius relation for ZAMS stars with solar metallicity (Baraffe et 
al. 1998 with fit parameters from Beuermann and Weichhold 1998, Eq. 5) provides a valid 
second relation between $M_2$ and $R_2$.
Combining both equations gives the mass-period relation 
\begin{equation}
\label{eq.m}
    M_2/M_{\sun} = 0.0686 \left({P_{\rm orb}/{\rm hours}}\right)^{1.59}\\
\end{equation}
from which we obtain $M_2 = 0.36\,M_{\sun}$ and $R_2 = 0.33\,R_{\sun}$ for the mass and 
radius of the secondary in RX\,J0501.7-0359, respectively. The corresponding mass-function
is shown in Fig.~\ref{masses}.

The duration of the eclipse, $\Delta T_{\rm ecl} = (13.1\pm 1.2)$\,min $= 
(0.077\pm 0.007)\,P_{\rm orb}$,
provides a second constraint for the inclination $i$ and the mass ratio $Q = M_1/M_2$ 
in this system (see Fig.~\ref{masses}). Combining both constraints the valid values for 
$i$ and $Q$ are restricted to the very narrow ranges $i = (75\pm 3)\deg$ and 
$Q = 1.20^{+0.18}_{-0.27}$ (shaded area in Fig.~\ref{masses}). Finally, using 
$M_{2}$ and $Q$ we get $M_{1} = 0.43^{+0.10}_{-0.07}\,M_{\sun}$ for the mass of the 
white dwarf.

Besides atomic emission lines, our optical spectra of RX\,J0513.7--0359 generally 
show a smooth continuum (top and middle panel in Fig.~\ref{ospec}). Only during one
observing run some hump structure might be present during the orbital phases 0.8--0.95 
(bottom panel in Fig.~\ref{ospec}). We have tentatively identified these features as the 
5th and 6th cyclotron harmonics corresponding to a magnetic field strength of 
$B\sim 25$\,MG in the accretion region. Given the weakness of the possible cyclotron 
signatures, the field strength of RX\,J0501.7-0359, however, has still to be regarded as 
uncertain.

The X-ray data show a very strong modulation with a bright phase lasting 
for about half of the orbital cycle (phases 0.35--0.75, cf. Fig.~\ref{licu}). 
This on-off modulation is most likely caused by the accretion pole 
disappearing behind the limb of the white dwarf.   
Two component (blackbody + thermal bremsstrahlung with 
$kT_{\rm tb} = 20\,$keV fixed) model fits to the ROSAT PSPC X-ray spectra 
which cover mainly the bright phase yield a blackbody temperature 
$kT_{\rm bb} = 38^{+5}_{-10}$\,eV  with an absorption column density of 
$N_{\rm H} = (0.78^{+0.14}_{-0.27})\,10^{21}$\,atoms/cm$^2$ which is 
slightly above the galactic value of $N_{\rm H,gal} = 0.60\,10^{21}$ 
(3$\sigma$ errors are given). 

The accretion geometry of RX\,J0501.7--0359 appears to be intriguing as 
the X-ray bright phase occurs around the superior conjunction of the white
dwarf. This implies that the visible accreting pole must be located on the
white dwarf hemisphere pointing away from the secondary star.

A more detailed analysis and discussion of our X-ray, optical, and IR data
of this interesting new eclipsing polar in the period gap will be presented
elsewhere (Burwitz et al., in prep.).


\acknowledgments

We thank the ROSAT team for its enduring work which resulted in the 
All-Sky-Survey data base and the staff at the La Silla and Calar Alto 
observatories for their competent assistance during our 
observing runs.
This work was supported by the DLR under grant 50\,OR\,9210\,5 
and, in part, by the Deutsche Forschungsgemeinschaft under grant Re 1100/3-1.  

%
%


\begin{references}
\reference Beuermann, K. 1998, High energy astronomy \& astrophysics, Agrawal P. (ed.), 
       Tata Inst.of Fund. Res., p. 100
\reference Beuermann, K., \& Burwitz, V. 
       1997, ASP Conf. Ser., 85, 99 
\reference Beuermann, K., Baraffe, I., Kolb, U., and Weichhold, M.
       1998, \aap, 339, 518
\reference Beuermann, K., Thomas, H.-C. 1990, \aap 230, 326
\reference Beuermann, K. \&  Weichhold, M.
       1998, these proceedings
\reference Burwitz, V., Reinsch, K., Beuermann, K., Thomas, H.-C. 1997, \aap, 327, 183 
\reference Burwitz, V., Reinsch, K., Schwope, A.D., et al.
        1998, \aap, 331, 262
\reference Thomas,~H.-C., Beuermann,~K., Reinsch,~K., Schwope,~A.D.,
       Tr\"umper, J., \& Voges, W. 1998, \aap, 335, 467
\reference Voges, W., Aschenbach, B., Boller, Th., et al. 1996, IAUC 6420,
       The ROSAT All-Sky Survey Bright Source Catalogue (1RXS), 
       http://www.rosat.mpe-garching.mpg.de/survey/rass-bsc/

\end{references}
\end{document}